\documentclass[9pt,conference]{IEEEtran}
\usepackage{dcase2026}

\usepackage{bm} %
\let\oldthebibliography\thebibliography
\renewcommand{\thebibliography}[1]{%
  \oldthebibliography{#1}%
  \normalsize%
}

\usepackage{dcase2026,amsmath,graphicx,url,times,booktabs, tabularx}

\usepackage{amssymb,pifont}
\newcommand{\cmark}{\ding{51}} %
\newcommand{\xmark}{\ding{55}} %

\title{Evaluating Inference-time Algorithms\\for Semantic Sound Scene Segmentation}

\name{Sripathi Sridhar$^{1,2}$,
      Gordon Wichern$^{2}$,
      Yoshiki Masuyama$^{2}$,
      Mark Cartwright$^{1}$,
      Jonathan Le Roux$^{2}$}
\address{$^{1}$New Jersey Institute of Technology, Newark NJ, USA \\
$^{2}$Mitsubishi Electric Research Laboratories (MERL), Cambridge, MA, USA\\
}

\begin{document}

\maketitle

\begin{abstract}
Spatial semantic segmentation of sound scenes (S5) involves the detection and extraction of target sound events in audio files. However, in typical detection-extraction pipelines, errors in the detection module propagate to the extraction module, degrading overall performance. In this work, we develop a multichannel detection-extraction model and evaluate inference-time algorithms to improve the class detection and target sound extraction in a multi-stage setup. We evaluate two fitness functions: binary cross entropy and mixture consistency; two relabeling strategies: naive and conditional confusion matrix relabeling; and three source estimate evaluation models: a multi-label classifier, a fine-tuned single-source classifier, and an off-the-shelf audio judge. We thoroughly evaluate the impact of these design choices on the DCASE 2025 Task 4 dataset, and validate entropy-based gating of inference-time updates on the held-out evaluation set, laying the groundwork for inference-time updates of S5 systems in real-world scenarios.
\end{abstract}


\section{Introduction}

Inference-time compute scaling, which focuses on improving the accuracy of responses by scaling computation at inference-time, has shown great success in LLM reasoning models \cite{snell_scaling_2024, setlur_e3_2025}. In audio, analogous gains have been pursued in applications such as source separation and speech enhancement, through the iterative refinement of signal representations \cite{bralios_latent_2023} or previously extracted target signals via multiple passes through the model \cite{manilow_improving_2022}.
Some models use iterative refinement during both training and inference \cite{manilow_improving_2022,ge_multi-stage_2021,deng_robust_2021}. Spex++ used extracted speech from a previous stage to condition subsequent stages \cite{ge_multi-stage_2021}, and in Iterative Refined Adaptation (IRA), the speaker embedding and speech latents iteratively updated and refined each other \cite{deng_robust_2021}. Others such as Zang et al.~\cite{zang_training-free_2025} adapted a single-step source separation model for multi-step inference by blending the mixture with the previous stage output in different ratios, improving performance on speech and music source separation. The underlying assumption though is that the target is well-defined, such as the speaker id or instrument classes in \textit{drums}, \textit{bass}, \textit{vocals}, and \textit{other}. However, this may not hold in detection-extraction systems where the target class must be detected first before extracting the target source.

In such a framework, errors from the detection module degrade target source extraction \cite{kwon2025sound}, and such errors cannot be fixed through iterative refinement in the source separation module alone. This challenge naturally arises in the DCASE 2025 Task 4, titled ``Spatial semantic segmentation of sound scenes (S5)''. Prior work on this task favored an iterative refinement approach in a chain-of-inference style, where previously extracted labels and sources condition subsequent source extraction steps, with some using the same models in consequent steps \cite{kwon_self-guided_2025, morocutti_temporal_2025} while others trained multiple classifiers and separators in a two-stage setup \cite{zhou_sjtu-audiocc_2025}.
While these methods refine the separated sources, they treat the predicted labels as reliable. Other approaches used label correction by comparing the predicted label for a source estimate against its input query \cite{park_performance_2025,Stergioulis_2025_t4}, or removed false positives and redid the source separation process when the number of predicted sources exceeded the maximum allowed \cite{park_performance_2025}. Concurrent submissions to the follow-up DCASE 2026 Task 4 explore inference-time candidate selection \cite{you_duplicate_2026} and re-tagging \cite{jeong_tagging_2026}. However, these approaches do not systematically explore more general label updates, fitness functions to score candidate labels, or label hypotheses beyond the model prediction. Evaluating these design choices could unlock a deeper understanding of inference-time algorithms for applications to general detection-extraction systems in real-world scenarios.

To this end, we propose inference-time algorithms to update predicted labels and the corresponding separated sources, and evaluate them on different detection models for the S5 task. We propose and evaluate different combinations of relabelers and fitness functions with a baseline, fine-tuned, and off-the-shelf detection model. We thoroughly analyze performance on the DCASE 2025 Task 4 test set and show improvements under certain conditions, laying the groundwork for inference-time algorithms in detection and separation systems. 

\section{Methods}
\subsection{Task and system description}

The S5 task \cite{yasuda2025description} in DCASE 2025 Challenge Task 4 requires systems to detect and separate individual sound events from multi-channel spatial audio mixtures. Each input mixture contains up to three simultaneous target sound events drawn from a vocabulary of $K=18$ classes, along with optional interference events and non-directional background noise. Systems must output isolated single-channel source estimates, each paired with a predicted class label. The primary evaluation metric is the class-aware signal-to-distortion ratio improvement (CA-SDRi) \cite{nguyen_baseline_2025}, which jointly rewards accurate source separation and correct class assignment by computing SDR improvement only when the predicted label matches the ground truth. Each false positive and false negative prediction is penalized with 0 SDRi.

Inspired by the baseline S5 system \cite{nguyen_baseline_2025}, we use a two-stage model consisting of an audio classifier followed by a target source extraction model. The classifier predicts active sound event labels from the input mixture, and these labels are used as prompts to query the extraction model, which produces one isolated source estimate per predicted label. We then apply a relabeling procedure that explores alternative label combinations and generate corresponding source estimates. Each alternative label combination is then ranked by making predictions for the corresponding source estimates and scoring them with a fitness function. We pick the alternative labels and resulting source estimates with the top rank as our final model output.

We use M2D \cite{niizumi_masked_2024}, a multi-label audio classifier model trained on Audioset \cite{gemmeke_audio_2017}, as our detection model. We use a publicly available checkpoint released by the challenge organizers \cite{nguyen_baseline_2025} that is fine-tuned on the DCASE 2025 Task 4 data. While other competitive audio classifiers exist such as BEATs \cite{chen_beats_2023}, EAT \cite{chen_eat_2024}, and SSLAM \cite{alex2025sslam}, we choose M2D since it performs competitively on AudioSet \cite{niizumi_masked_2024} while allowing us to build on the baseline system and compare directly to challenge submissions. Moreover, although the magnitude of performance gains may vary across different classifiers, the pipeline itself remains classifier-agnostic, since it relies only on the classifier's own probabilities and confusion matrix.

\subsection{Target source extraction model}

We build our target source extraction model based on the task-aware unified source separation (TUSS) architecture \cite{saijo_task-aware_2025}. Unlike language-queried models such as AudioSep \cite{liu_separate_2023} that extract one source per text query, TUSS uses context from all the provided prompts to jointly extract target sources from the input mixture, which suits the closed-set S5 task. We refer the reader to \cite{saijo_task-aware_2025} for further details. We base our model on the publicly available code\footnote{\url{https://github.com/merlresearch/unified-source-separation}}.

TUSS uses a fixed prompt vocabulary during training and inference, but can extract a flexible number of sources during inference. We use the 18 target sound classes in DCASE 2025 Task 4 \cite{nguyen_baseline_2025} as the prompt vocabulary, and add separate non-target `interference' and `noise' prompts, where the interference prompt is repeated to match the number of interference sources in the training example. At inference, the number of interference sources is unknown, whereas background noise is always present during training and inference. We develop a multichannel TUSS variant, Mc-TUSS, to take advantage of the four-channel input audio in this task.

We train Mc-TUSS from scratch on the DCASE 2025 Task 4 training data on a single GPU using batch size two for 150 epochs, with other training parameters following TUSS \cite{saijo_task-aware_2025}. We evaluate the checkpoint with the best validation loss in all inference-time experiments. Mc-TUSS matches or outperforms non-iterative variants of the top two ranked submissions \cite{kwon_self-guided_2025, morocutti_temporal_2025}, achieving 15.1 dB SNRi with ground-truth labels on the test set. 

\subsection{Relabelers}
We explore different algorithms to update the predicted labels that prompt the target sound extraction model. In this work, `relabeling' refers to the selection of alternate labels different from the classifier predicted labels. Initial experiments with directly sampling from the sigmoid logits of the multilabel classifier prediction did not yield promising results due to the model's tendency to confuse certain classes with each other. Per-class decision thresholds could also be optimized \cite{lipton_optimal_2014}, but do not exploit the structure of inter-class confusions. Thus, we explore two relabeling strategies based on the classifier's confusion matrix. In addition to alternate labels for target events, we also relabel the number of interference sources from the set \{0, 1, 2\} during target+non-target relabeling.

\subsubsection{Na\"ive Confusion Matrix (CM) relabeling}
We compute the confusion matrix from single-source examples and select the top 5 confusions per class. Given a set of predicted labels, we enumerate the combinations of all possible replacements for each predicted label as our set of alternative labels. Since there are a maximum of three target events per example, the maximum cardinality of the search space is 15. 

\subsubsection{Conditional Confusion Matrix (CCM) relabeling}
The na\"ive confusion matrix relabeler preserves the number of predicted sources, which limits its ability to correct errors where M2D predicts the wrong number of active classes. We propose to address this limitation by combining the audio classifier probabilities with the confusion matrix, producing corrected class probabilities over all 18 classes that can shift probability mass to classes not originally predicted.
\label{subsec:ccm}

Let the baseline M2D sigmoid output vector be $\mathbf{p} \in [0,1]^K$ over $K=18$ classes, and define $\mathbf{C} \in \mathbb{R}^{K \times K}$ as the column-normalized confusion matrix, where $C_{ij} = P(\text{true} = i \mid \text{predicted} = j)$. We compute corrected class probabilities as 
\begin{equation}
    \tilde{\mathbf{p}} = \mathbf{C} \mathbf{p},
\end{equation}
clipped to $[0, 1]$. Each sigmoid probability $p_j$ is redistributed according to the confusion pattern for class $j$. High-confidence predictions contribute proportionally more, and frequently confused classes receive probability mass from the classes they are mistaken for. Crucially, the corrected vector $\tilde{\mathbf{p}}$ can assign nonzero probability to classes absent from the original M2D prediction, enabling the relabeler to explore combinations with different numbers of active sources. This resembles the forward correction for label noise \cite{patrini_making_2017}, applied here at inference time to generate label hypotheses rather than to correct a training loss.

We score candidate label multihot vectors $\mathbf{u} \in \{0,1\}^K$ by their log-probability
\begin{equation}
    \log P(\mathbf{u} \mid \tilde{\mathbf{p}}) = \sum_{i=1}^{K} \left[ u_i \log \tilde{p}_i + (1 - u_i) \log (1 - \tilde{p}_i) \right].\label{eq:logp}
\end{equation}
Since the task constrains the number of active sources to between one and three, the set of valid class combinations has size $\binom{18}{1} + \binom{18}{2} + \binom{18}{3} = 987$. We enumerate all 987 combinations, score each using Eq.~\cref{eq:logp}, sort by descending log-probability, and retain the top $N$ candidates for evaluation by the fitness function. We set $N=100$ unless specified otherwise. This deterministic enumeration avoids wasteful sampling and guarantees that no candidate is evaluated more than once. 

\subsection{Source estimate evaluation models}
We evaluate several evaluation models that generate probability vectors or ratings for the source estimates generated from relabeled prompts. The outputs of the evaluation models are used by the fitness functions to rank the alternative labels. We evaluate M2D, which is used to detect the initial labels, also as an evaluation model.

While M2D performs well on mixtures, and even on reference single-class sources, we found that M2D exhibits more inter-class confusions when evaluated on Mc-TUSS source estimates from ground-truth labels as queries. We address this by finetuning M2D with Mc-TUSS source estimates using a cosine classifier head and ArcFace loss \cite{deng_arcface_2019}. We extract source estimates from Mc-TUSS for 100k examples from the training set, select the source estimates with SNRi $\geq$ 3dB, and finetune just the classifier head while keeping the rest of the model weights fixed. Similar to the winning team of DCASE 2025 Task 4 \cite{kwon_self-guided_2025}, we trained for 100 epochs with learning rate 1e-4 and use validation loss to pick the best checkpoint. We refer to this model trained on single-class source estimates as M2D-SC.

We also evaluate SAM Audio Judge \cite{wang_sam_2026}, a judge model trained on human ratings of source separation quality in speech, music, and environmental sound data. Given a text prompt, input mixture, and the separated source, SAM Audio Judge rates the separated source along dimensions of precision, recall, faithfulness, and overall quality. 

\subsection{Fitness functions}

We evaluate three fitness functions to score the quality of source estimates conditioned on alternate labels provided by the relabeler. For target sources, we use binary cross-entropy (BCE) when the evaluation model is M2D or M2D-SC, or a specific dimension (e.g., recall) when using SAM Audio Judge. We use mixture consistency to optimize the number of non-target prompts (interference and noise). 

Let $\hat{\mathbf{u}} \in \{0,1\}^K$ be a candidate relabeled multi-hot vector with corresponding class set $\hat{\mathcal{T}} = \{t : \hat{u}_t = 1\}$, and $\hat{x}_t$  the separated source estimate for a target class $t$. Let $\hat{\textbf{p}}_t \in [0,1]^K$ denote the sigmoid output of the evaluation model applied to the separated source estimate $\hat{x}_t$, and $\mathbf{v} \in \{0,1\}^K$ the initial M2D multi-hot prediction from the mixture.

\subsubsection{Binary Cross-Entropy (BCE)} 
 For M2D, we assume that confusion patterns remain consistent between the mixture and the source estimates:
\begin{equation}
    \mathcal{L}_{\text{M2D}}(\hat{\mathbf{u}}) = -\sum_{t \in \hat{\mathcal{T}}} \sum_{k=1}^{K} \left[ v_k \log \hat{p}_{tk} + (1 - v_k) \log (1 - \hat{p}_{tk}) \right]
    \label{eq:bce}
\end{equation}
For M2D-SC, which may not retain those confusion patterns, we compute BCE against the relabeled multi-hot vector $\hat{\mathbf{u}}$ instead, replacing $v_k$ with $\hat{u}_k$ in Eq.~\cref{eq:bce}.
In both cases, we select the relabeled classes that minimize $\mathcal{L}(\hat{\textbf{u}})$. 

\subsubsection{Mixture Consistency}
Mixture consistency is defined as the mean squared error (MSE) between the mono input mixture $y$ and the sum of all target ($\hat{x}_t$) and non-target ($\hat{x}_{\text{n}}$) source estimates, $\mathcal{L}_{\text{mc}} = \text{MSE}\left( y, \sum_t \hat{x}_t + \sum_n \hat{x}_{\text{n}} \right)$.
When jointly evaluating BCE and mixture consistency, we scale the mixture consistency loss by 100, following the weighting strategy used during joint classification-separation training in prior work \cite{pishdadian_learning_2020}. We pick the sample that minimizes $\mathcal{L}_{\text{bce}} + 100 \mathcal{L}_{\text{mc}}$.

\subsubsection{SAM Audio Judge score}
SAM Audio Judge (SAJ) \cite{wang_sam_2026} predicts a score $r$ given an input mixture, a text prompt, and a source estimate. We use the relabeled classes as text prompts and compute the mean score across all estimates within a clip:
\begin{equation}
    \bar{r}(\hat{\mathbf{u}}) = \frac{1}{|\hat{\mathcal{T}}|} \sum_{t \in \hat{\mathcal{T}}} r(y,\hat{x}_t, t)
\end{equation}
We define a joint score $S(\hat{\mathbf{u}}) = \bar{r}(\hat{\mathbf{u}}) - \alpha \, \mathcal{L}_{\text{mc}}(\hat{\mathbf{u}})$ combining SAJ with mixture consistency, and select the candidate that improves over the baseline:
\begin{equation}
    \hat{\mathbf{u}}^* = \begin{cases} \displaystyle\arg\max_{\hat{\mathbf{u}}} S(\hat{\mathbf{u}}) & \text{if } \displaystyle\max_{\hat{\mathbf{u}}} S(\hat{\mathbf{u}}) > S(\mathbf{v}) \\ \mathbf{v} & \text{otherwise} \end{cases}
\end{equation}
When using SAJ alone, we set $\alpha=0$. When combining with mixture consistency, we set $\alpha=1000$.

\subsection{Gating functions for inference-time algorithms}
\label{subsec:gating-methods}
Inference-time updates may not always be useful when applied indiscriminately, for instance on examples with high baseline CA-SDRi. We explore algorithmic approaches to automate this decision without access to ground-truth sources or labels. Specifically, we investigate SAJ scores and M2D binary entropy.

We compute M2D binary entropy for each class $k$ of the $K=18$ classes: $H_k = -p_k\log p_k - (1-p_k)\log (1-p_k)$
and aggregate to clip-level features $H_\text{max}=\max_k(H_k)$ , $H_\text{sum}=\sum_{k=1}^K H_k$, and $H_\text{norm}=H_\text{sum}/n_{active}$, where $n_{active}$ is the number of active targets.

We sweep 500 thresholds and evaluate performance on the test set for the SAJ scores and the binary entropy features. We pick an oracle threshold with maximum combined $\Delta$CA-SDRi over gated and ungated examples, where gated examples are the ones inference-time updates are applied to. We assess generalization by applying them unchanged to the held-out DCASE 2025 Task 4 evaluation set containing 1620 examples.
\begin{figure}[t]
    \centering
    \includegraphics[width=\columnwidth]{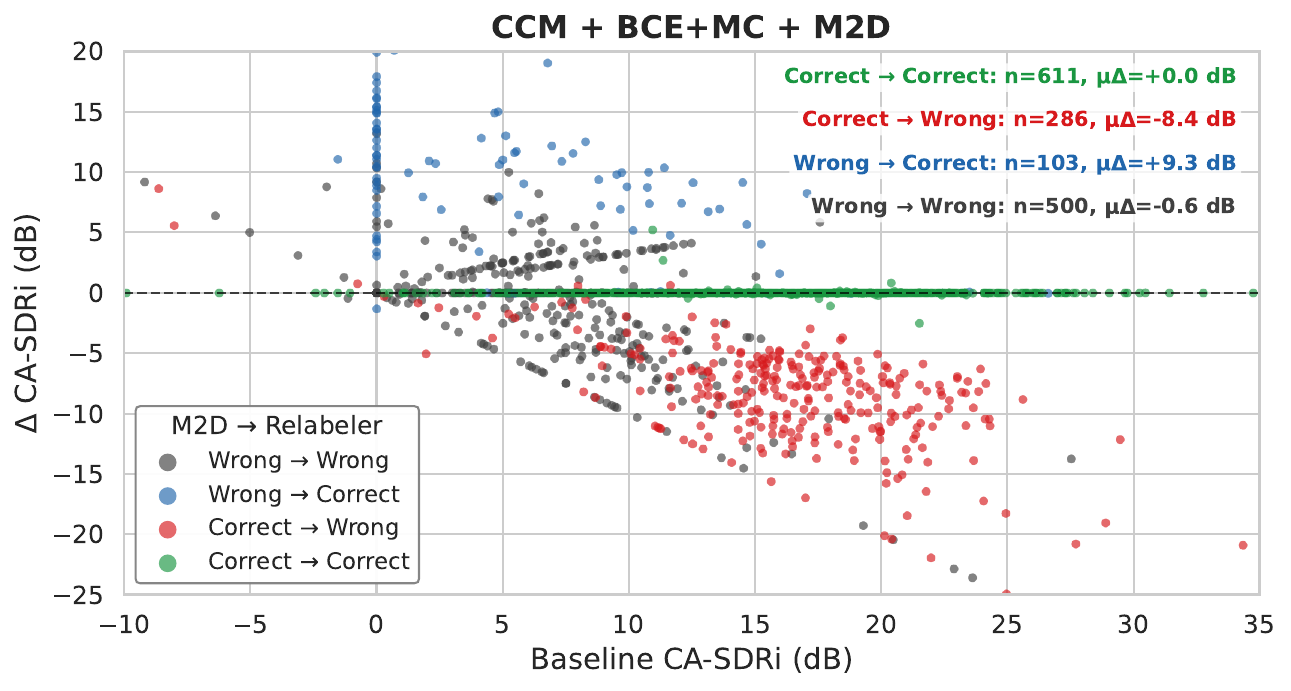}
    \caption{Delta CA-SDRi vs.\ baseline CA-SDRi on the full test set for M2D with conditional CM relabeler and BCE+MC fitness function.}
    \label{fig:delta_vs_baseline}
\end{figure}

\section{Results and Discussion}

We evaluate all combinations of fitness function, relabeler, and evaluation models on the test set. We use the relabeled classes as the class predictions for the metric. We report results on the full test set, and subsets with examples that match certain conditions, in \cref{tab:relabeler-results}.

\subsection{Performance breakdown on the full test set}

Our algorithms do not improve performance when applied indiscriminately. This trend holds across relabeler, fitness, and evaluation model types. We note that the relabeler does present useful hypotheses for evaluation, as evidenced by the positive $\Delta_\text{oracle}$ values for all relabeler configurations, which evaluates relabeled classes by computing CA-SDRi against the ground truth. Moreover, the conditional confusion matrix relabeler, which can alter the number of sources in the relabeled classes, presents better hypotheses, as seen in larger $\Delta_\text{oracle}$ values. This suggests that the fitness function is the bottleneck.

To investigate this further, we qualitatively analyze the best performing model, M2D with CCM relabeling and BCE+MC fitness function. We depict separation improvements as a function of the baseline performance, CA-SDRi by using the initial M2D prompts, in \cref{fig:delta_vs_baseline}. We notice that the model generally improves performance on examples with poor baseline performance, e.g., CA-SDRi $\leq$ 10dB, either by correcting the baseline predictions to the correct labels (%
blue dots, Wrong $\rightarrow$ Correct), or updating them to other wrong labels (grey dots, Wrong $\rightarrow$ Wrong) but which better isolate the true positive target sources. Based on this observation, we analyze the performance of our algorithms on specific subsets of the test set where maximum gains may be achieved.

\begin{table*}[t]
\centering
\caption{Performance of various prompt relabeling strategies, fitness functions, and evaluation models on the DCASE 2025 Task 4 test set. CM: Na\"ive Confusion matrix, CCM: Conditional Confusion matrix, BCE: Binary cross entropy, MC: Mixture consistency, M2D: Multilabel classifier, M2D-SC: Single-class classifier, SAJ: SAM Audio Judge. $\Delta$ refers to improvement over the baseline CA-SDRi, while Oracle is the theoretical maximum improvement from the hypotheses evaluated given an ideal fitness function, i.e., computing CA-SDRi against ground truth. Bold indicates best performance; underline indicates second best.}
\vspace{+.3cm}
\label{tab:relabeler-results}
\setlength{\tabcolsep}{4pt}
\renewcommand{\arraystretch}{1.1}
\begin{tabular}{c c c c c c c c c c c c c}
\hline
& & & & \multicolumn{3}{c}{\textbf{All (1500 ex.)}} & \multicolumn{3}{c}{\textbf{$\leq$10 dB (563 ex.)}} & \multicolumn{3}{c}{\textbf{Oracle Gap (51 ex.)}} \\
& & & & \multicolumn{3}{c}{Baseline: 12.24 dB} & \multicolumn{3}{c}{Baseline: 4.79 dB} & \multicolumn{3}{c}{Baseline: $-$0.21 dB} \\
\cmidrule(lr){5-7} \cmidrule(lr){8-10} \cmidrule(lr){11-13}
\textbf{Relabeler} & \textbf{Non-Target?} & \textbf{Fitness} & \textbf{Evaluator} & \textbf{$\Delta$} & \textbf{CA-SDRi} & \textbf{$\Delta_\text{oracle}$} & \textbf{$\Delta$} & \textbf{CA-SDRi} & \textbf{$\Delta_\text{oracle}$} & \textbf{$\Delta$} & \textbf{CA-SDRi} & \textbf{$\Delta_\text{oracle}$} \\
\hline
CM & \xmark & BCE & M2D & -4.10 & 8.15 & 1.24 & -0.43 & 4.37 & 3.08 & 2.61 & 2.40 & 10.04 \\
CM & \cmark & BCE+MC & M2D & -4.02 & 8.22 & 1.42 & -0.28 & 4.52 & 3.49 & 2.06 & 1.85 & 10.35 \\
CM & \xmark & BCE & M2D-SC & -2.68 & 9.56 & 1.24 & 0.47 & 5.26 & 3.08 & 5.34 & 5.13 & 10.04 \\
CM & \cmark & BCE+MC & M2D-SC & -2.48 & 9.77 & 1.42 & \underline{0.84} & \underline{5.64} & 3.49 & 5.22 & 5.01 & 10.35 \\
CM & \xmark & Recall & SAJ & -2.45 & 9.79 & 1.24 & 0.24 & 5.03 & 3.08 & 4.01 & 3.80 & 10.04 \\
CM & \cmark & Recall+MC & SAJ & \underline{-1.90} & \underline{10.35} & 1.42 & 0.67 & 5.47 & 3.49 & 4.52 & 4.31 & 10.35 \\
\hline
CCM & \xmark & BCE & M2D & -3.85 & 8.40 & 3.48 & 0.78 & 5.57 & 7.71 & 4.83 & 4.62 & 14.84 \\
CCM & \cmark & BCE+MC & M2D & \textbf{-1.16} & \textbf{11.08} & 3.48 & \textbf{1.56} & \textbf{6.36} & 7.71 & 4.85 & 4.64 & 14.84 \\
CCM & \xmark & BCE & M2D-SC & -4.39 & 7.86 & 3.48 & 0.42 & 5.21 & 7.71 & \textbf{5.43} & \textbf{5.22} & 14.84 \\
CCM & \cmark & BCE+MC & M2D-SC & -3.78 & 8.46 & 3.48 & 0.42 & 5.22 & 7.71 & \underline{5.40} & \underline{5.19} & 14.84 \\
CCM & \xmark & Recall & SAJ & -7.82 & 4.43 & 3.52 & -1.89 & 2.91 & 7.72 & 2.03 & 1.82 & 14.84 \\
CCM & \cmark & Recall+MC & SAJ & -7.10 & 5.15 & 3.52 & -1.43 & 3.36 & 7.72 & 2.49 & 2.21 & 14.84 \\
\hline
\end{tabular}
\end{table*}

\subsection{Performance breakdown on specific subsets}
In \cref{tab:relabeler-results}, we present results for examples in the test set with baseline CA-SDRi $\leq$ 10 dB. We also show results on oracle gap examples, which have baseline CA-SDRi $<$ 0dB but oracle CA-SDRi $>$ 5 dB, i.e., where finding the correct labels is likely to improve performance.
M2D is best overall, and improves CA-SDRi by 1.56 dB on the $\leq10$~dB examples, which constitutes more than a third of the test set. M2D-SC matches or beats SAJ across relabeler types. M2D-SC improves performance over M2D when using the naive CM relabeler, but seems incompatible with CCM as seen by its degraded performance. This is likely due to the addition of false positives and negatives by CCM, leading to penalties and big swings in CA-SDRi.

Combining mixture consistency with other fitness functions such as BCE and SAJ recall improves performance. Mixture consistency adds a useful mixture constraint on the relabeler, and adding non-target prompts can reduce leakage into target source estimates. We also find that the choice of relabeler is conditional on the evaluation model. While the CCM relabeler significantly improves performance for M2D over the naive CM, especially in the combined BCE+MC fitness function case, it degrades performance for M2D-SC and SAJ. 

However, it is clear that the proposed algorithms are useful for examples with poor baseline CA-SDRi performance. While we manually selected these examples by computing CA-SDRi against ground-truth sources here, we view this oracle analysis as a first step towards human-in-the-loop iterative inference, where a listener flags clips with unsatisfactory separation for inference-time updates. An alternate approach is to automatically identify potential candidates for inference-time updates to develop a more end-to-end system.

\subsection{Automatically detecting candidate examples for inference-time updates}
We present gating results for our best performing model so far, i.e., M2D with conditional confusion matrix relabeling on target and non-target prompts using BCE+MC as the fitness function, in \cref{tab:gating,tab:gating-eval}. Together with $\Delta_\text{oracle}$ in \cref{tab:relabeler-results}, which bounds the improvement given an ideal fitness function, \cref{tab:gating} provides upper bounds for gating functions since their thresholds are oracle-selected on the test set itself.

On the test set, M2D entropy features provide a better gating signal. SAJ score selects examples with already strong baseline performance; while there are some performance gains on the gated examples, the relative improvement is less significant and the support set is also small.
On the other hand, H$_\text{sum}$ conservatively picks 24 examples with low baseline CA-SDRi of 2.85 dB and boosts performance by 1.96 dB on those examples, a relative improvement of 69\%. H$_\text{norm}$ is less conservative, picking 165 examples ($\approx$10\% of the test set) but improving performance by a smaller margin. This suggests a tradeoff between candidate selection and performance gains which could be decided based on the application scenario.

\cref{tab:gating-eval} shows what the frozen thresholds achieve in practice on the held-out evaluation set. Similar to \cref{fig:delta_vs_baseline}, SAJ gates select clips with baseline CA-SDRi near or above 10 dB, where updates are likely to be less useful. The entropy gates on the other hand select low-baseline CA-SDRi clips. However only H$_\text{norm}$ remains useful, gating 298 clips ($\approx$18\% of the eval set) with 4.92 dB baseline and improving them by +0.18 dB (+0.032 dB overall). Though small, these gains suggest that entropy-based gating could identify suitable candidates on unseen data without re-tuning gating thresholds, with headroom available towards the oracle bounds presented in Table \ref{tab:gating}.

\begin{table}[t]
\centering
\caption{Gating threshold sweep results on the test set. SAM Audio Judge scores (R=Recall, P=Precision, O=Overall quality, F=Faithfulness) gate clips with score $\leq \tau$; M2D entropy features gate clips with score $> \tau$. Thresholds $\tau$ are oracle-selected by the sweep. $N$: number of gated clips. $\Delta$: change in overall CA-SDRi vs.\ baseline (12.24~dB). $\Delta_{\text{g}}$: mean CA-SDRi change for gated clips. $\text{BL}_{\text{g}}$, $\text{BL}_{\text{r}}$: mean baseline CA-SDRi for gated and remaining clips.}
\vspace{+.3cm}
\label{tab:gating}
\setlength{\tabcolsep}{3pt}
\begin{tabular}{lccccccc}
\toprule
Feature & $\tau$ & $N$ & CA-SDRi & $\Delta\uparrow$ & $\Delta_{\text{g}}\uparrow$ & ${\text{BL}}_{\text{g}}\downarrow$ & ${\text{BL}}_{\text{r}}\uparrow$ \\
\midrule
SAJ$_R$ & 0.97 & 5 & 12.25 & +0.003 & +0.90 & 12.21 & 12.24 \\
SAJ$_P$ & 1.77 & 57 & 12.28 & +0.031 & +0.82 & 9.68 & 12.35 \\
SAJ$_O$ & 0.99 & 17 & 12.25 & +0.004 & +0.35 & 11.62 & 12.25 \\
SAJ$_F$ & 1.01 & 55 & 12.27 & +0.024 & +0.67 & 12.04 & 12.25 \\
\hline
H$_\text{sum}$ & 3.20 & 24 & 12.28 & +0.031 & \textbf{+1.96} & \textbf{2.85} & 12.40 \\
H$_\text{max}$ & 0.69 & 89 & 12.27 & +0.023 & +0.39 & 8.54 & 12.48 \\
H$_\text{norm}$ & 1.61 & 165 & \textbf{12.29} & \textbf{+0.046} & +0.42 & 6.22 & \textbf{12.99} \\
\bottomrule
\end{tabular}
\end{table}

\begin{table}[t]
\centering
\caption{Gating on the held-out DCASE 2025 Task 4 evaluation set (1620 clips) with thresholds $\tau$ frozen from \cref{tab:gating}. $\Delta$: change in overall CA-SDRi vs.\ baseline (8.61~dB); other columns as in \cref{tab:gating}.}
\vspace{+0.3cm}
\label{tab:gating-eval}
\setlength{\tabcolsep}{3pt}
\begin{tabular}{lccccccc}
\toprule
Feature & $\tau$ & $N$ & CA-SDRi & $\Delta\uparrow$ & $\Delta_{\text{g}}\uparrow$ & ${\text{BL}}_{\text{g}}\downarrow$ & ${\text{BL}}_{\text{r}}\uparrow$ \\
\midrule
SAJ$_R$ & 0.97 & 4 & 8.61 & 0.000 & $-$0.10 & 11.59 & 8.60 \\
SAJ$_P$ & 1.77 & 81 & 8.50 & $-$0.106 & $-$2.13 & 10.52 & 8.51 \\
SAJ$_O$ & 0.99 & 35 & 8.58 & $-$0.030 & $-$1.41 & 9.56 & 8.59 \\
SAJ$_F$ & 1.01 & 73 & 8.55 & $-$0.057 & $-$1.27 & 10.69 & 8.51 \\
\hline
H$_\text{sum}$ & 3.20 & 37 & 8.60 & $-$0.005 & $-$0.24 & \textbf{4.62} & 8.70 \\
H$_\text{max}$ & 0.69 & 95 & 8.54 & $-$0.070 & $-$1.20 & 7.58 & 8.67 \\
H$_\text{norm}$ & 1.61 & 298 & \textbf{8.64} & \textbf{+0.032} & \textbf{+0.18} & 4.92 & \textbf{9.44} \\
\bottomrule
\end{tabular}
\end{table}

\section{Conclusion}

In this work, we systematically evaluate inference-time updates for the S5 task in a classifier-separator pipeline. We study different relabelers, fitness functions, and source estimate evaluation models. We find that our best inference-time algorithms improve performance on examples where the separation model performs poorly to begin with. Applied indiscriminately, however, these updates degrade overall performance. We explore automatic gating functions that select examples likely to benefit from our algorithms. Entropy-based gating achieves a small positive gain on the held-out evaluation set without threshold re-tuning. Our oracle analyses point to higher upper bounds, which leaves headroom for stronger fitness and gating functions. Overall, our work charts the design space of these inference-time algorithms and lays a foundation for their use in detection-separation systems. A limitation of our setup is that we do not evaluate on examples with more than one source per target class or no target sources, like in DCASE 2026 Task 4 \cite{yasuda_description_2026}. This could be addressed for instance by source-counting  \cite{saijo_merl_2026} or relaxing the label hypothesis space, which we leave to future work.

\clearpage
\bibliographystyle{IEEEtran}
\bibliography{refs}

\end{document}